\documentclass[]{spie}  

\usepackage{amsmath,amsfonts,amssymb}
\usepackage{aas_macros}
\usepackage{graphicx}
\usepackage[colorlinks=true, allcolors=blue]{hyperref}
\usepackage{xcolor}

\title{The Pandora SmallSat: post-launch operations and mission update}
\author[*a, b]{Trevor O. Foote}
\author[a]{Elisa V. Quintana}
\author[c]{Jessie L. Dotson}
\author[a]{Knicole D. Col\'on}
\author[a]{Thomas Barclay}

\author[d]{Pete Supsinskas}
\author[d]{Jordan Karburn}

\author[e,f,g]{D\'aniel Apai}
\author[a,h]{Christina Hedges}
\author[i]{Benjamin V. Rackham}
\author[j]{Jason F. Rowe}
\author[k]{David Ciardi}
\author[k]{Emily A. Gilbert}

\author[l]{Natalie H. Allen}
\author[k]{Jessie Christiansen}
\author[m,l]{N\'estor Espinoza}
\author[f]{Nguyen Fuda}
\author[n]{Aylin Garcia Soto}
\author[k]{Thomas P. Greene}
\author[j]{Kelsey Hoffman}
\author[a,b]{Rae Holcomb}
\author[a,b]{Benjamin J. Hord}
\author[a,b]{Aishwarya R. Iyer}
\author[k]{Aurora Y. Kesseli}
\author[a,o]{Veselin B. Kostov}
\author[p]{Nikole K. Lewis}
\author[q]{Sarah E. Logsdon}
\author[r]{Andrew W. Mann}
\author[s]{Megan Mansfield}
\author[t]{James Mason}
\author[d]{Peter McGill}
\author[a,w]{Julie Moquin}
\author[m]{Brett M. Morris}
\author[a]{Gregory Mosby}
\author[m]{Susan E. Mullally}
\author[n]{Elisabeth R. Newton}
\author[a,s]{Joe P. Renaud}
\author[v]{Yoav Rotman}
\author[f]{Suman Saha}
\author[a]{Joshua E. Schlieder}
\author[v]{Luis Welbanks}
\author[t]{Lindsey S. Wiser}
\author[a]{Allison Youngblood}
\author[a]{Robert T. Zellem}

\affil[a]{NASA Goddard Space Flight Center, Greenbelt, MD 20771, USA}
\affil[b]{NASA Postdoctoral Program Fellow}
\affil[c]{NASA Ames Research Center, Moffett Field, CA 94035, USA}
\affil[d]{Lawrence Livermore National Laboratory, Livermore, CA 94550, USA}
\affil[e]{Steward Observatory, The University of Arizona, Tucson, AZ 85721, USA}
\affil[f]{Lunar and Planetary Laboratory, The University of Arizona, Tucson, AZ 85721, USA}
\affil[g]{James C. Wyant College of Optical Sciences, The University of Arizona, AZ 85721, USA}
\affil[h]{University of Maryland, Baltimore County, Baltimore, MD 21250, USA}
\affil[i]{Department of Earth, Atmospheric and Planetary Sciences, and Kavli Institute 
for Astrophysics and Space Research, Massachusetts Institute of Technology, Cambridge, MA 02139, USA}
\affil[j]{Department of Physics and Astronomy, Bishops University, 2600 Rue College, Sherbrooke, QC J1M 1Z7, Canada }
\affil[k]{Caltech/IPAC-NASA Exoplanet Science Institute, Pasadena, CA 91125, USA}
\affil[l]{Department of Physics and Astronomy, Johns Hopkins University, Baltimore, MD 21218, USA}
\affil[m]{Space Telescope Science Institute, 3700 San Martin Drive, Baltimore, MD 21218, USA}
\affil[n]{Department of Physics and Astronomy, Dartmouth College, Hanover NH 03755, USA}
\affil[o]{SETI Institute, 189 Bernardo Ave, Suite 200, Mountain View, CA 94043, USA}
\affil[p]{Department of Astronomy, Cornell University, Ithaca, NY 14853, USA}
\affil[q]{NSF NOIRLab, 950 N.\ Cherry Ave., Tucson, AZ 85719, USA}
\affil[r]{Department of Physics and Astronomy, The University of North Carolina at Chapel Hill, Chapel Hill, NC 27599, USA}
\affil[s]{University of Maryland, College Park, College Park, MD 20742 USA}
\affil[t]{Johns Hopkins University Applied Physics Laboratory, Laurel, MD 20723, USA}
\affil[u]{Dartmouth College, Hanover, NH 03755, USA}
\affil[v]{School of Earth and Space Exploration, Arizona State University, 781 Terrace Mall, Tempe, AZ, 85287, USA}
\affil[w]{Southeastern Universities Research Association, Washington, DC 20005, USA}

\authorinfo{* trevor.o.foote@nasa.gov}

\begin{document} 
\maketitle

\begin{abstract}
The Pandora SmallSat is a NASA Pioneers mission designed to disentangle stellar activity from exoplanetary atmospheric signatures in transmission spectroscopy observations by simultaneously performing visible photometry and near-infrared spectroscopy. Pandora’s baseline mission lifetime planned for a one-month commissioning period, followed by a 12-month prime mission period for science observations. Herein, we present a comprehensive post-launch mission update, detailing operational performance and current observational strategy. We will also highlight key operational milestones, challenges encountered during commissioning and early science operations, and solutions implemented. This information provides insight into Pandora's on-orbit capabilities for precision multi-wavelength observations that enable reliable determination of exoplanet atmosphere compositions.

\end{abstract}

\section{INTRODUCTION}
\label{sec:intro} 

To date, transmission spectroscopy has proven to be one of the most powerful techniques available for the characterization of exoplanetary atmospheres. With the operational success of the James Webb Space Telescope (JWST), the exoplanet community has increasingly focused on the “small star opportunity.” This approach leverages the favorable planet-to-star size ratio inherent to M-dwarf and late K-dwarf systems to probe the atmospheres of sub-Neptune and Earth-sized planets \cite{Nutzman2008, Triaud2021}. However, these lower-mass host stars are generally more active than Sun-like stars. The presence of surface inhomogeneities, such as starspots and their bright counterparts, faculae, presents a major challenge for exoplanet transmission spectroscopy \cite{Rackham2018, Rackham2019}. When an exoplanet transits a spotted star, the unocculted stellar surface spectrum can differ significantly from the transit chord. This causes stellar spectral features to be imprinted onto the planetary transmission spectrum. Ultimately, this stellar contamination can mimic or suppress true atmospheric features in the intrinsic planetary spectrum.

The Pandora SmallSat, selected under the NASA Astrophysics Pioneers Program, is designed to directly address this challenge by disentangling stellar and planetary signals. Successfully launched on January 11, 2026, into a Sun-synchronous low Earth polar orbit, Pandora employs a dual-channel observational approach. The observatory simultaneously captures visible photometry ($\sim$0.4--0.7 $\mu$m) to measure time-varying stellar brightness and near-infrared (NIR) spectroscopy ($\sim$0.8--1.6 $\mu$m) to probe planetary atmospheric composition along with stellar spectral variations. By combining these datasets, the mission aims to provide the robust data necessary to accurately identify water- or hydrogen-dominated atmospheres.

Pandora’s nominal mission architecture outlined a one-month commissioning period followed by a 12-month prime science phase. However, the transition from ground-testing to in-orbit operations presented unique environmental and operational complexities. Consequently, Pandora has undergone an extended commissioning phase. This additional time was necessary to fully characterize the flight performance of the detector assemblies and refine the spacecraft's observational parameters.

Herein, we present a comprehensive post-launch mission update detailing Pandora’s early on-orbit performance. We highlight key operational milestones achieved since the January 2026 launch and discuss the specific challenges encountered during the extended commissioning. We also outline the operational solutions implemented by the team and detail the mission's current observational strategy. This provides valuable insight into Pandora’s in-flight capabilities for delivering the precision multi-wavelength observations required to reliably characterize exoplanet atmospheres.

\section{SCIENCE OBJECTIVES}

The primary goal of the Pandora mission is to disentangle stellar and planetary signals within transmission spectra. This enables the reliable determination of exoplanetary atmospheric compositions. To achieve this goal, the mission architecture and science requirements are driven by two principal science objectives.

The first objective focuses on determining the spot and faculae covering fractions of low-mass exoplanet host stars and assessing the impact of these active regions on exoplanetary transmission spectra. To accomplish this, Pandora utilizes a visible channel to acquire precise time-series photometric measurements of a star's brightness. Concurrently, a NIR channel operates in tandem with the visible channel to capture variations in spectral features that correlate with the observed stellar photometric variability.

By combining long-baseline visible photometry with simultaneous NIR spectroscopy, Pandora provides the comprehensive spectral information required to decompose the stellar spectrum into its constituent parts, namely, spots, faculae, and the quiescent photosphere \cite{Zhang2018, Wakeford2019, Narrett2024}. This simultaneous multi-wavelength approach allows for the measurement of key physical quantities, such as spot contrasts and covering fractions, precisely at the time of a planetary transit. Extracting these parameters is essential, as it enables the calculation of robust upper limits on stellar contamination signals \cite{Rackham2018, Wakeford2019, Garcia2022, Rackham2023}.

Building upon the stellar contamination mitigations developed under the first objective, Pandora’s second objective is to robustly identify exoplanets possessing hydrogen- or water-dominated atmospheres. It will also determine which of these targets are likely covered by clouds and hazes. To fulfill this objective, Pandora will obtain transmission spectra across a wavelength bandpass that encompasses strong molecular water absorption features, which are prominent in both hydrogen- and water-dominated atmospheric regimes.

\section{MISSION OVERVIEW}

On January 11, 2026, Pandora successfully launched from Vandenberg Space Force Base. During insertion the observatory was placed into a Sun-synchronous low Earth orbit with an altitude of approximately 605\,km. This allows Pandora to observe the entirety of the sky over the course of 12 months of science operations.

Pandora's baseline mission architecture requires the observation of at least 10 transits for each of its 20 primary exoplanet targets \cite{Colon2025}. A typical observation, defined as a visit, consists of a continuous 24-hour window encompassing the planetary transit event. The transit is not required to be perfectly centered within this temporal window but does require a minimum of 4 hours pre- and post-transit baseline. Due to the geometry of the low Earth orbit, the primary target is periodically occulted by the Earth for approximately 20 minutes of each 100-minute orbit. During these occultation periods, the observatory slews to an alternate astronomical target to maximize observing efficiency. Once the primary target emerges from the Earth's limb, the spacecraft slews back to resume the primary science sequence.

While the observatory remains power-positive during the majority of its science operations, supplemental battery power is required during seasonal orbital eclipses and when observing targets located near the solar exclusion zone. Pointing near this boundary results in sub-optimal solar array incidence angles, temporarily creating a negative power balance. Furthermore, although most observations utilize a fixed inertial pointing strategy, specific thermal constraints must be managed during the eclipse season. For certain target geometries, the spacecraft must execute a 180-degree rotation about the boresight every half-orbit to ensure proper thermal dissipation and maintain the stability of the detector assemblies.

Science and telemetry data are downlinked multiple times per day utilizing both X-band and S-band communication. Because downlinking requires the spacecraft to actively slew the X-band antenna toward a KSAT ground station, these communication windows are preferentially scheduled during Earth occultations to avoid interrupting primary science observations. However, specific orbital geometries occasionally necessitate downlinks that temporarily interrupt the primary target visit. To mitigate this, Pandora features onboard storage that significantly exceeds the daily downlink capacity. This allows the operations team to execute high-data-rate observations and temporarily buffer the data for transmission during periods with lower collection rates.

The ground segment architecture is divided into distinct operational and data processing centers. The Mission Operations Center (MOC) at the University of Arizona is responsible for generating spacecraft commands, scheduling ground contacts, transmitting command loads, and staging the downlinked telemetry. The MOC formulates slewing commands based on target schedules delivered by the Science Operations Center (SOC) based primarily at NASA Goddard Space Flight Center, actively avoiding downlink interruptions during critical planetary transit windows. Once staged by the MOC, science data are transferred to the Data Processing Center (DPC) at NASA Ames Research Center. The DPC ingests the raw telemetry into the Pandora analysis pipeline to generate higher-level data products, including calibrated images, photometric light curves, and transmission spectra. These finalized products are subsequently delivered to the NASA Exoplanet Archive on a regular cadence for public access.

\section{THE OBSERVATORY}

\begin{figure}
    \centering
    \includegraphics[width=0.75\linewidth]{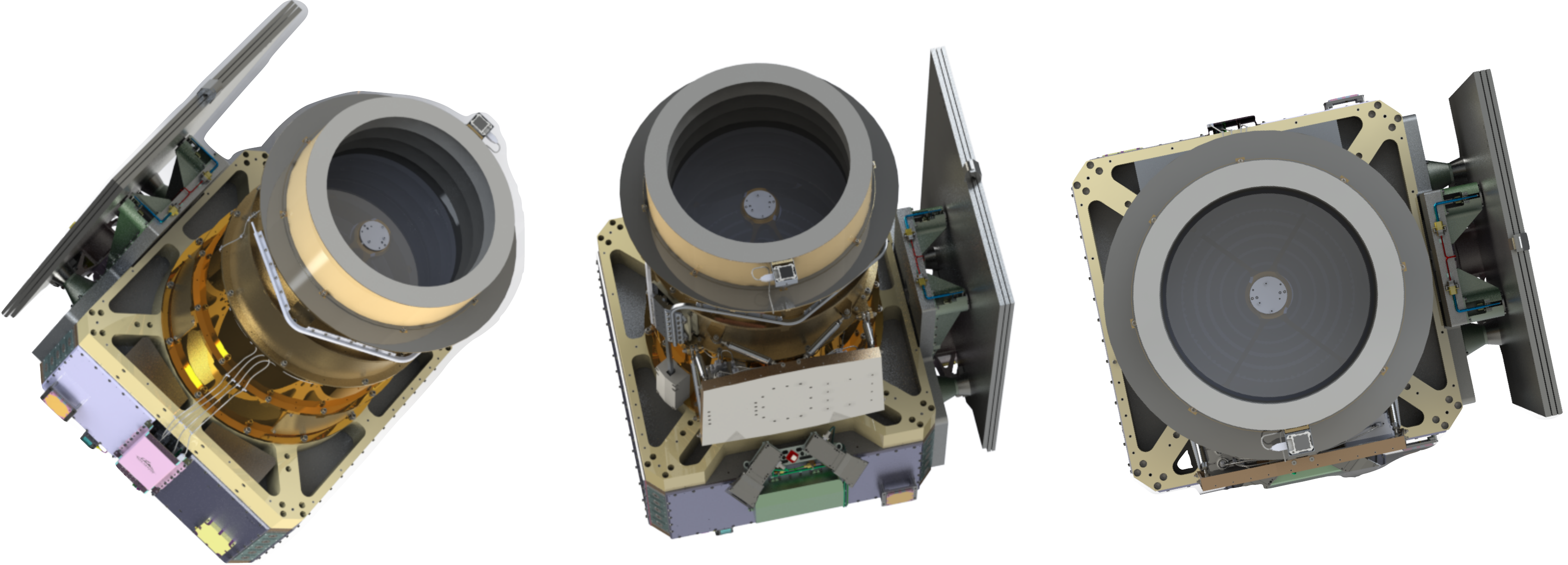}
    \caption{The Pandora Observatory in stowed configuration is designed to meet the ESPA Grande volume limits. Three renders show the observatory from different angles with the solar array stowed. The spacecraft bus encompasses the lower section of the observatory, with the payload attached to the bus via an interface plate.}
    \label{fig:observatory}
\end{figure}

The observatory itself is comprised of two major segments: a Blue Canyon Technologies (BCT) Saturn bus, and the payload developed by Lawrence Livermore National Laboratory (LLNL). Figure~\ref{fig:observatory} illustrates the observatory in its stowed configuration. 

The Blue Canyon Technologies spacecraft bus provides the critical infrastructure required to support Pandora's scientific payload during its on-orbit operations. This segment is responsible for managing essential observatory functions, including power generation, thermal regulation, and telemetry communication with ground stations. Additionally, the bus houses the attitude determination and control system, which delivers the precise pointing stability necessary to conduct long-baseline transmission spectroscopy of exoplanet transits.

The scientific payload houses the dual-channel telescope assembly designed to simultaneously capture the visible and near-infrared observations required to disentangle stellar contamination from exoplanet transmission spectra. Following the light path, the payload's first component is the 45-cm Cassegrain telescope with an outer baffle assembly designed to provide off-axis light suppression. As the light exits the telescope, it is directed into the optical relay assembly, where a dichroic mirror splits the NIR and visible wavelengths and shepherds them to their respective detector assemblies. For the visible light, it is imaged directly onto the visible detector, while the NIR light is first dispersed using a prism, then captured onto the NIR detector. Unseen by the light path is the cooling system used to actively cool the detectors as well as the Payload Electronics Module (PEM), which controls the payload.

\subsection{Detector Assemblies}

\begin{figure}
    \centering
    \includegraphics[width=0.75\linewidth]{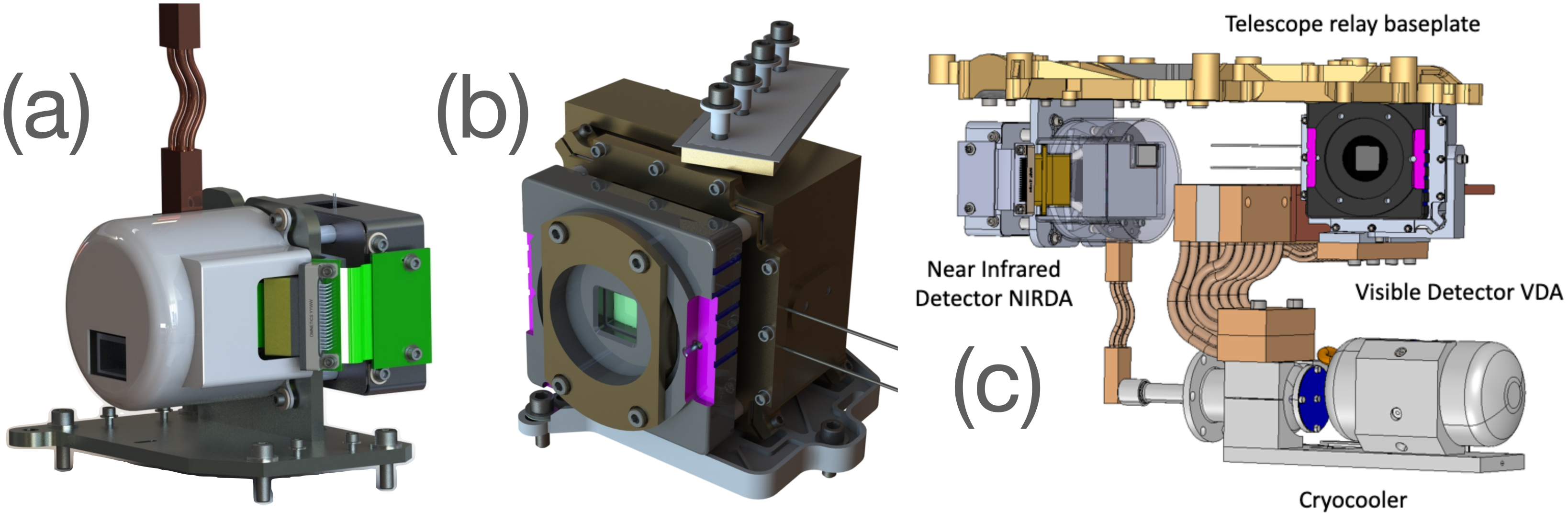}
    \caption{The dual detector assemblies enable Pandora to operate the two channels simultaneously. Panel (a) shows the NIRDA which consists of a Teledyne H2RG sensor and SIDECAR ASIC, and panel (b) shows the VDA which includes an Excelitas pco.panda 4.2 CMOS sensor. The mechanical layout of the two detectors assemblies is shown in panel (c) along with the cryocooler which is used to maintain thermal control and stability.}
    \label{fig:nidra-vda}
\end{figure}

The Visible Detector Assembly (VDA) is composed of an Excelitas pco.panda 4.2 CMOS detector and a thermoelectric cooler. A rendering of it is provided in Figure~\ref{fig:nidra-vda}b. The size of the CMOS detector is a 2048$\times$2048 pixel array with a pixel scale of 0.78"/pixel. However, there is a field stop within the focal plane to reduce stray light which reduces the area of illuminated pixels to roughly the center 1600$\times$1600 pixels. The VDA's wavelength sensitivity is from 400--750 nm. During operations the VDA's native framerate is at 5~Hz. 

Images from the VDA serve two primary goals on Pandora. The first is to augment Pandora's pointing stability, which is described in further detail in Sec.~\ref{sec:pointing}. For this goal the native 5~Hz frames are utilized and all analysis is completed onboard the spacecraft. The second is to collect our science images of the target star as well as nearby bright stars ($V_{mag} \leq 16.5$) which are used as photometric references. For these observations we need to downlink the images to the ground for processing by the DPC and provided to the community for analysis. To efficiently transfer this data to the ground and reduce the data volume that needs to be downlinked, a couple of preliminary processing steps occur onboard the spacecraft. Following the successful completion of an observation, the 5~Hz frames are coadded by 50 frames resulting in an effective integration time of 10~s. The other way Pandora reduces the total data downlinked from the VDA is rather than downlinking the entire array of pixels, it extracts between 1--9 50$\times$50 pixel regions of interest, typically centered on stars in the field of view.

The Near Infra-Red Detector Assembly (NIRDA) consists of a Teledyne HAWAII-2RG (H2RG) sensor and SIDECAR Application Specific Integrated Circuit (ASIC), enclosed within a custom mechanical housing. Because both the sensor and ASIC are flight spares from the JWST Near Infrared Camera (NIRCam) instrument \cite{Rieke2005}, Pandora benefits substantially from the rigorous pre-flight testing conducted by the NIRCam team \cite{Schlawin2020,Schlawin2021,Schlawin2023}. Although the H2RG detector possesses a native 2.5$\mu$m cutoff, the optical path includes a red-edge filter that rejects wavelengths longer than 1.63$\mu$m to mitigate excess thermal background. Additionally, a blue-edge filter blocks wavelengths shorter than 0.8$\mu$m. This restricts the spatial length of the observed spectra, preventing signal contamination from adjacent stars within the field of view. Figure~\ref{fig:nidra-vda}a illustrates the NIRDA, highlighting the rectangular optical entrance aperture on the left-hand side. Data digitized by the SIDECAR ASIC are subsequently read by a Markury Scientific MACIE card before being transmitted to the payload electronics. To maintain thermal stability, the sensor is actively cooled by a dedicated cryocooler and associated control electronics, depicted in Figure~\ref{fig:nidra-vda}c.

While the H2RG sensor consists of a 2048$\times$2048 pixel array, Pandora restricts science observations to a small sub-section of the detector. An aluminum outer shield and an internal snout enclose the majority of the sensor to block stray light and thermal noise originating from inside the housing. This configuration leaves an exposed active region of approximately 1200$\times$600 pixels situated near the sensor edge. To mitigate the impact of 1/f noise on the spectroscopic data, stellar spectra are specifically dispersed along the slow read direction of the detector \cite{Schlawin2020, Schlawin2021}. An individual spectrum spans approximately 200 pixels, allowing the mission to operate the sensor in a highly efficient 400$\times$80 pixel subarray mode. This targeted readout geometry yields a rapid frame time of 0.32s. A standard integration consists of 24 sequential frames, resulting in a total integration time of 7.68s. Integrations are collected continuously while the observatory maintains pointing on a science target. To conserve onboard data storage and downlink bandwidth, NIRDA observations employ a Fowler-4 sampling strategy. In this scheme, the first four frames are averaged, the subsequent 16 frames are skipped, and the final four frames are averaged to complete the integration.

\section{OPERATIONAL PERFORMANCE}

\subsection{Pointing stability} \label{sec:pointing}
Achieving sub-arcsecond pointing stability is critical to meeting Pandora’s photometric and spectroscopic performance requirements. Maintaining precise pointing over timescales of tens of minutes significantly mitigates systematic noise sources, such as inter- and intra-pixel sensitivity variations and spatial flat-field anomalies on the detector. While the native pointing capabilities of the BCT Saturn-class bus provide excellent baseline stability, with a nominal performance of $\sim$1 to 2$^{\prime\prime}$, we require sub-arcsecond stability for our high-precision science goals.

To bridge this performance gap and achieve sub-arcsecond stability, LLNL developed a custom fine-pointing architecture known as Visible In the Loop (VITL). This system leverages Pandora's visible science detector to act as a highly sensitive fine guidance sensor. During standard operations, initial target acquisition is executed using the default attitude determination and control system (ADCS) of the Saturn bus. Once the spacecraft has successfully slewed to the desired coordinates, the payload electronics transition into the VITL operational mode to refine the pointing.

Within the VITL framework, pointing updates are continuously calculated by centroiding the brightest stars ($V_{mag} \leq 16.5$) across the visible science array. These calculations are performed at the native camera readout rate of 5~Hz. The payload electronics then interfaces with the bus' ADCS, sending the calculated attitude errors and requesting their correction within the ADCS to maintain the target position. 

A portion of Pandora's extended commissioning phase was dedicated to tuning the VITL correction algorithm and identifying the optimal bus settings for applying these corrections. Furthermore, the team developed best practices regarding the precise timing for activating the VITL system following a slew to a new target. We determined that successful activation requires accounting for two distinct settling periods. First, the team must consider the mechanical settling time of the solar array following slew maneuvers. Second, sufficient time must be allocated for the default ADCS pointing algorithm to stabilize. Following these on-orbit refinements, Pandora now routinely achieves sub-arcsecond pointing stability during science operations.

\subsection{NIRDA Thermal Control and Stability}

The thermal performance of the near-infrared detector assembly (NIRDA) is a primary factor that directly impacts the quality of Pandora's spectroscopic science data. There are two key thermal metrics for the NIRDA that dictate performance: the absolute operating temperature and the temperature stability over time. Detector systematics within the HAWAII-2RG (H2RG) sensor, such as the prevalence of bad pixels and image persistence, are highly temperature-dependent and are significantly mitigated at cooler operating temperatures \cite{Beletic2008}. During early operations, the mission team has demonstrated the ability to routinely cool and maintain the NIR detector at an absolute temperature of 130~K. While data calibration is achievable at this baseline temperature, correcting for these temperature-dependent effects remains a significant challenge. Consequently, the operations team is actively exploring new thermal management strategies to drive the detector to cooler operating temperatures, targeting temperatures closer to those tested during ground testing (90--110~K) \cite{Foote2024}.

Equally critical to the absolute baseline temperature is the high-precision thermal stability of the NIRDA. The electronic performance of the NIR detector is highly sensitive to thermal fluctuations over the duration of an exoplanet transit observation. By maintaining a stable thermal environment, the team can develop effective noise models for eliminating or reducing temperature-dependent noise sources like dark current and 1/f noise. Throughout routine operations, Pandora has successfully achieved the required thermal stability for the NIR channel. Specifically, the NIR detector temperature is consistently maintained to within $<$ 30~mK.

\subsection{VDA Image Quality}

Prior to launch, potential degradation of the optical system due to the launch environment was a tracked mission risk. On-orbit data confirm the telescope assembly did not experience structural degradation, and initial test images exceeded baseline image quality expectations. However, analysis of the point spread function (PSF) reveals variations in the spatial profile that correlate strongly with the temperature differential between the primary and secondary mirrors. Although the PSF remains within science requirements under the worst-case observed thermal conditions, the operations team is actively investigating thermal control strategies to minimize this differential and stabilize the PSF across observations.

\subsection{Observatory Pointing Constraints}

During both active observations and slewing maneuvers, the operations team enforces strict pointing constraints that have evolved significantly since the beginning of commissioning. Prior to launch, these constraints focused primarily on protecting the sensitive detector assemblies and preserving data quality from scattered light. Based on pre-launch optical modeling, the observatory boresight was assigned minimum exclusion angles of $90^\circ$ from the Sun, $20^\circ$ from the Earth limb, and $25^\circ$ from the Moon \cite{Foote2023}.

The first major modification to these operational constraints was driven by on-orbit stray light characterization. Analysis of early science data revealed a significant increase in background flux when the telescope boresight is oriented within $45^\circ$ of the illuminated Earth limb. To address this contamination, the team implemented a scheduling mitigation strategy utilizing the mission planning software \cite{Foote2023}. By updating the scheduling algorithms to enforce a stricter $45^\circ$ Earth-limb avoidance constraint, the observatory successfully minimizes stray light during critical transit observations. 

Beyond modifying existing optical constraints, the extended commissioning phase revealed the necessity for new pointing constraints related to the spacecraft star trackers. These star trackers dictate the default ADCS fine guidance, which must stabilize the spacecraft before the VITL architecture can initialize. On-orbit testing demonstrated that the star trackers are more sensitive to Earth-shine and scattered light than originally anticipated. Specifically, the trackers fail to yield valid attitude solutions when their individual boresights violate exclusion angles of $30^\circ$ from the Earth limb, $50^\circ$ from the Sun, or $20^\circ$ from the Moon. If both star trackers fail prior to VITL initialization, the VITL sequence aborts, resulting in unacceptable pointing stability.

Fortunately, operational testing identified several robust mitigating factors regarding star tracker performance. The team discovered that only one of the two star trackers requires a valid solution to achieve the baseline stability necessary to successfully engage the VITL algorithm. Furthermore, once VITL is actively controlling the pointing, the system exhibits significant tolerance to temporary star tracker outages. Under current operating conditions, the observatory can maintain sub-arcsecond pointing precision through continuous star tracker dropouts lasting up to 12 minutes before pointing degradation occurs.

\section{EARLY-CAREER RESEARCHERS' INVOLVEMENT}
early-career
A secondary objective of both the NASA Astrophysics Pioneers program and the Pandora mission is to provide substantive space mission experience to early-career participants. Throughout the mission development phase, the Pandora team deliberately incorporated early-career scientists and engineers into critical roles. The framework and early accomplishments of this mentorship approach were previously documented by \cite{Dotson2024}. Following the successful launch and transition into on-orbit operations, the mission has continued to prioritize the integration of early-career researchers into highly impactful operational roles.


While senior team members provide continuous guidance and support, the daily operations within the SOC are predominantly executed by these early career researchers. This arrangement provides an exceptional opportunity for participants to gain comprehensive expertise in space flight operations. In addition to executing routine tasks such as generating observing schedules, these researchers actively coordinate observatory activities with the Mission Operations Center. Furthermore, they are responsible for identifying spacecraft anomalies, formulating mitigation strategies, and generating formal status reports for the NASA Program Office. Ultimately, this approach ensures the next generation of scientists acquires extensive, hands-on experience in mission operations.

\section{SUMMARY}

The Pandora SmallSat mission, successfully launched in January 2026, represents a critical step forward in the multi-wavelength characterization of exoplanetary atmospheres. By simultaneously capturing visible photometry and near-infrared spectroscopy, the observatory is uniquely positioned to disentangle stellar activity from planetary transmission spectra. This dual-channel capability allows the science team to derive spot and faculae covering fractions, enabling the robust identification of water- and hydrogen-dominated atmospheres around low-mass stars.

Transitioning from ground testing to active low Earth orbit operations required an extended commissioning phase to fully characterize and optimize the spacecraft subsystems. Through dedicated on-orbit tuning, the mission has successfully demonstrated its required performance capabilities, with the custom Visible In the Loop architecture routinely delivering sub-arcsecond pointing stability. Furthermore, the near-infrared detector assembly maintains a highly stable baseline temperature of 130~K, and the optical telescope assembly continues to exceed pre-launch image quality expectations.

To maximize the scientific yield of the observatory, the operations team has continuously refined Pandora's observing schedules and spacecraft pointing constraints based on empirical flight data. Stricter Earth-limb avoidance angles and dynamic star tracker management strategies have been successfully implemented to mitigate stray light and maintain continuous attitude control. As the team finalizes the remaining commissioning tasks, these robust operational procedures will ensure a seamless transition into the primary science operations phase. Ultimately, the resulting calibrated data products and transmission spectra will provide the exoplanet community with unprecedented insights into planetary atmospheric compositions in the presence of stellar activity.

\acknowledgments 
Pandora is supported by NASA's Astrophysics Pioneers Program. Pre-proposal work was supported by GSFC's Internal Research and Development Program and a Mission Planning Lab at Wallops Flight Facility. We are indebted to the many people who contributed to the various iterations of Pandora, and to the technical team at LLNL who have been key to the success of the mission. TOF, BJH, AI, and RH would like to acknowledge support from the NASA Postdoctoral Program by an appointment at the NASA Goddard Space Flight Center, administered by Oak Ridge Associated Universities under contract with NASA.

\bibliography{pandora} 
\bibliographystyle{spiebib} 

\end{document}